\documentclass[10pt,a4]{article}
\usepackage{graphics}

\begin{document}
%
%
\large
\begin{center}
{\bf Dynamics of a rod in a random static environment:}
{\bf non-Gaussianity at large length scales}
\end{center}
\normalsize
\begin{center}
A.J. MORENO and W. KOB
\end{center}
\small
\begin{center}
Laboratoire des Verres. Universit\'{e} de Montpellier II.
Place E. Bataillon. CC 069. F-34095 Montpellier, France.
\end{center}
\normalsize
\begin{center}
ABSTRACT
\end{center}
\small
\hspace{0.5 cm}We present molecular dynamics simulations of the motion of a single rod in a
two-dimensional random static array of disks. For long rods the mean-squared displacement of
the center-of-mass shows a cage effect similar to that observed in supercooled liquids
or dense colloidal systems. We have determined the time-dependence of the non-Gaussian parameter
for different rod lengths. It is found that the long-time regime is strongly non-Gaussian
even at length scales of the order of 10-15 times the rod length, thus showing the heterogeneity
of the dynamics at such length scales.
\\
\normalsize
\begin{center}
\S 1 INTRODUCTION
\end{center}
\hspace{0.5 cm} 
The Lorentz gas is a simple model in which a particle diffuses in a random configuration of 
static obstacles. Initially introduced by Lorentz as a model for the electrical conductivity in metals
(Lorentz 1905), it can be used as a rough approximation for the dynamics of
a light atom in a disordered environment which presents a much slower dynamics.
It has been shown that, for high densities of obstacles, it qualitatively shows many of the features of
the dynamics of supercooled liquids or dense colloidal systems, such as a transition from an ergodic
phase of non-zero diffusivity to a non-ergodic phase with no diffusivity 
(Bruin 1972, Alder and Alley 1978, G\"{o}tze {\it et al.} 1981a,b, Masters and Keyes 1982). 
Despite its simplicity, this problem is highly non-trivial even in the simplest case
where the diffusing particle and the obstacles are modelled as hard-spheres. In particular,
diffusion constants and correlation functions are non-analytical functions of the density of obstacles
(Bruin 1972, G\"{o}tze {\it et al.} 1981a,b, Binder and Frenkel 1990).
In this work we present an investigation by means of molecular dynamics simulations on
a generalization of the Lorentz gas: the diffusion of a rod in a
random configuration of static obstacles. This picture can be understood as a cartoon of the dynamics
of a linear molecule in a porous medium or in a colloidal suspension of heavy particles.
Though simulations have been done in two dimensions, we do not expect qualitatively different features
in the general three-dimensional case.

The paper is organized as follows: In Section 2 we present the model and give some details of
the simulation. General dynamical features are shown in Section 3. A study on the non-Gaussianity
of the long-time regime is presented in Section 4. Conclusions are given in Section 5.  
\newline
\begin{center}
\S 2 MODEL AND DETAILS OF THE SIMULATION
\end{center}
\hspace{0.5 cm} The rod consists of $N$ beads of mass $m$ and radius $\sigma$ equal to that of the obstacles.
The rod length is therefore $2N\sigma$.
The density of obstacles is defined as $\rho = N_{\rm obs}/L_{\rm box}^{2}$, with $N_{\rm obs}$ the number of
obstacles and $L_{\rm box}$ the length of the square simulation box.
Beads and obstacles interact via a soft-sphere potential $V(r) = \epsilon(\sigma/r)^{12}$. The
latter is truncated and shifted at a cutoff distance of 2.5$\sigma$. In the following, space and time
will be measured in the reduced units $\sigma$ and $(\sigma^{2}m/\epsilon)^{1/2}$ respectively.

A set of 100 different realizations of the rod was propagated in a given configuration of the obstacles.
Periodic boundary conditions were imposed.
The set was equilibrated at temperature $T = \epsilon/k_{\rm B}$ by coupling it to a stochastic heat bath.
After the equilibration, a run of $10^{8}$ time steps, with a step size of 0.01,
was performed at constant energy. In order to improve the statistics, 
this procedure was repeated for 6-10 different configurations of the obstacles
and the obtained results were averaged over the corresponding runs. 
\newline
\begin{center}
\S 3. TRAJECTORIES AND DYNAMICS
\end{center}
\hspace{0.5 cm} In the following, all the results will be presented for a fixed density
of obstacles $\rho = 6\times 10^{-3}$.
An extensive investigation for a wide range of $N$ and $\rho$ will be presented elsewhere.
Figure 1 shows typical trajectories for two
characteristic rod lengths $N =$ 5 and 50. They respectively correspond to rods of dimensions
comparable and much longer than the average distance between nearest-neighbour obstacles,
$d_{\rm n.n.} \approx 7$. For $N = 5$, the trajectory presents 
a structure of blobs and jagged channels. Thus, the rods are able to penetrate often in regions
of high local density of obstacles, where they are trapped for some time, forming the blobs, while
the jagged channels correspond to paths in regions of lower local density, where the collisions
are less frequent.

A very different polygonal-like structure is obtained for $N = 50$, consisting of nearly 
longitudinal motions in the tubes formed by the obstacles, connected by vertexes corresponding to
regions where the rod leaves the tube and enters into a new one.

Figure 2a shows the mean-squared displacement of
the center-of-mass of the rod $\langle r^{2}(t)\rangle$ divided by the time $t$. Brackets denote
ensemble average and $r(t)= |{\bf R}(t)-{\bf R}(0)|$, with ${\bf R}(t)$ the position
of the center-of-mass of the rod at time $t$. At short times the rod does not
feel the presence of the obstacles and $\langle r^{2}(t)\rangle$ shows a quadratic time-dependence,
as expected for a ballistic motion. For small $N$ a sharp transition to the long-time
regime $\langle r^{2}(t)\rangle \sim t$
is observed, while for long rods a crossover regime,
which extends over 1-2 decades, develops at intermediate times. This corresponds to the well-known
cage effect observed in supercooled liquids (Kob and Andersen 1995, Sciortino {\it et al.} 1996,
Caprion and Schober 2000, Mossa {\it el al.} 2000, Colmenero {\it et al.} 2002) 
or dense colloidal systems (Puertas {\it et al.} 2002, van Megen 2002, Weeks and Weitz 2002): the particle is 
temporarily trapped by the cage formed by their neighbours and as a consequence,
the collisions within the cage result in a time-dependence of the mean-squared displacement 
that is slower than the one in the ballistic regime. Finally, the particle escapes
from the cage and begins to diffuse. It is worthwhile to remark that in the present case
the breaking of the cage is not achieved by cooperative motions between neighbouring particles,
as in the case of liquids or colloids, since the configuration of obstacles is static.
Instead, the trapping of the rod is of purely entropic nature. 
It must also be noticed that the observed cage effect is much less
pronounced than in supercooled liquids or dense colloids. 
Higher densities of obstacles or longer rods would be required for observing a stronger effect.
\newline
\begin{center}
\S 4. NON-GAUSSIANITY
\end{center}
\hspace{0.5 cm} In Einstein's random walk model,
particles move under the effect of collisions with the others. Between two consecutive collisions,
the particle moves along a straight line in a random direction that is independent of its previous history.
If the time and distance between two consecutive collisions are
much smaller respectively than the observational time and lenght scales, the van Hove self-correlation
function takes a Gaussian form (see for example Hansen and McDonald 1986).
This result is exact for an ideal gas and for an harmonic crystal.
It is also valid in the limit $t \rightarrow 0$, where atoms
behave as free particles.  
At time and length scales comparable to those of the dynamics of the collisions, i.e., at observational
scales sensitive to heterogeneities of the system, deviations from Gaussianity are expected.
They can be quantified by the second-order non-Gaussian
parameter $\alpha_{2}(t)$, which for a two-dimensional motion
is given by $\alpha_{2}(t) = [\langle r^{4}(t)\rangle/2\langle r^{2}(t)\rangle^{2}]-1$.
It is straightforward to see that for a Gaussian function $\alpha_{2}(t)=0$. 

Figure 2b shows the $N$-dependence of the non-Gaussian parameter.
While for short rods a broad structure is obtained, for long ones a peak develops, which grows and shifts
to longer times for increasing $N$. A qualitatively similar behaviour has been reported for 
other very different systems as a function of different control parameters, as temperature for
supercooled liquids (Kob and Andersen 1995, Sciortino {\it et al.} 1996, Caprion and Shober 2000, Mossa {\it et al.} 2000)
or volume fraction for colloid-polymer systems 
(Puertas {\it et al.} 2002). However, in contrast to these systems, we find that the peak does
not grow in a completely monotonic way, and that for short rods ($N = 5, 10$ in figure 2b), it seems to be located 
at {\it longer} times that for longer rods. 

From a comparison with figure 2a, it is deduced that for long rods
($N > 10$) the time $t^{\ast}$ for the maximum of $\alpha_{2}(t)$ is located at the time interval connecting 
the end of the cage regime and the beginning of the long-time linear regime, i.e., the breaking of the cage
leads to a high probability of jumps much longer that the size of the cage, resulting in a strongly
non-Gaussian -or heterogeneous- dynamics at that time scale. Similar results have been reported for other systems
(Kob and Andersen 1995, Sciortino {\it et al.} 1996, Caprion and Shober 2000, Mossa {\it et al.} 2000,
Colmenero {\it et al.} 2002). The size of the cage $r_{\rm c}$ can approximately be estimated as
$r_{\rm c} = \langle r^{2}(t^{\ast})\rangle^{1/2}$.
We find a value of $r_{\rm c} \approx 35$, independent of $N$ within the error bar.
This value is notably larger than the typical distance between neighbouring obstacles $d_{\rm n.n.}\approx 7$. 
However, it must be noted that the latter perhaps is not the most appropiate length for
characterizing the caging dynamics of the rod trapped by the neighbouring obstacles. 
In contrast to a liquid-like
equilibrium configuration, which shows an approximately homogeneus structure,
a random configuration is characterized by clusters of close particles and big holes.
For the density $\rho = 6\times 10^{-3}$, the size of the holes, $r_{\rm h}$,
is typically between 40 and 70, which introduces another characteristic length.
The size of the cage should lie at some value between $d_{\rm n.n.}$ and $r_{\rm h}$, 
consistent with the estimated value of $r_{\rm c}$.

In the case of short rods ($N \leq 10$), where no intermediate cage regime is observed, the maximum of $\alpha_{2}(t)$
does not seem to be located at the sharp transition between the ballistic and the long-time regime,
but at a later time.
The reason for this difference with the large-$N$ case might be that, as argued in Section 3,
for these lengths the rod has
a high probability of exploring the holes, and at the same time is able
to penetrate in regions of high local density, leading to non-Gaussian dynamics at length scales of the order
of the hole size.          

The most striking feature in figure 2b is that in the time limit of the simulation, $\alpha_{2}(t)$ has not
decayed to zero but remains finite, especially for long rods, i.e., the long-time regime is still non-Gaussian
at the length scale covered by the simulation. This effect can be studied by investigating the $q$-dependence
of the incoherent intermediate scattering function
$F_{\rm s}(q,t)=\langle\exp(-i{\bf q}\cdot[{\bf R}(t)-{\bf R}(0)])\rangle$, where brackets denote ensemble
and angular average.
 
Figure 3 shows $F_{\rm s}(q,t)$ for $N = 50$ and several values of the wavelength $\lambda = 2\pi/q$
corresponding to length scales much larger than the size of the rod.
A two step-decay separating the cage and long-time regimes is not observable for this range of $\lambda$,
as in the case of supercooled liquids, where the plateau in $F_{\rm s}(q,t)$ is only observable
at $q$-values not too far from the position of the first maximum of the structure factor.
 
In Gaussian approximation in two dimensions, the incoherent intermediate scattering function is given by
$F_{\rm s}(q,t) = \exp(-\langle r^{2}(t)\rangle q^{2}/4)$.
Therefore, Gaussianity in the long-time  regime is only
fullfilled if the time dependence of the long-time decay of $F_{\rm s}(q,t)$ can be well described by an
exponential function $\exp(-t/\tau(q))$, since at that time scale $\langle r^{2}(t)\rangle \sim t$.
Figure 3a shows the corresponding fits to exponential functions. It must be stressed that, usually,  an amplitude
factor $A(q)<1$ must be included in the fit to take into account the strength of the long-time regime.
While in other cases this is an important parameter, for the present case
setting $A(q)=1$ did not change the results from the fit within the error bar, since the
value of $F_{\rm s}(q,t)$ is very close to unity at $t \sim 10^{3}$, the time for the
beginning of the long-time regime (see figure 2a). From figure 3a it is clear that only for wavelengths $\lambda$ 
beyond $\approx$ 2000 -i.e., about 20 times the rod length- the decay of the curve can be reasonably
described in terms of an exponential function,
and even for such large scales non-Gaussianity cannot be excluded, since the final decay is not accessed within
the time window of the simulation.

A better description can be achieved (figure 3b)  by fitting to a Kohlrausch-Williams-Watts (KWW) function,
$\exp[-(t/\tau(q))^\beta]$, a phenomenological equation often used in the analysis
of long-time relaxations of complex systems (Phillips 1996). It must be noted that these type of fits are not fully satisfactory:
as can be seen in figure 3b, the final part of the long-time decay relaxes slower than a KWW function.
In any case, the obtained values of $\beta$, different from unity (see inset in figure 3b),
again confirm the non-Gaussian nature of the long-time linear regime at large length scales.
\newline
\begin{center}
\S 5. CONCLUSIONS
\end{center}
\hspace{0.5 cm}
We have carried out molecular dynamics simulations for a simple model of the motion of a rod in a
random static environment.
The dynamics of long rods show similar features to those observed in supercooled liquids and dense colloids,
namely a cage effect at intermediate times between the ballistic and long-time regimes, due to the trapping
of the rod inside the tube formed by the neighbouring obstacles.

The time-dependence of the non-Gaussian parameter has been calculated for different rod lengths. It is found that
it does not decay to zero within the time window of the simulation, which for long rods cover length scales
of the order of 10-15 times the rod length. Therefore, the dynamics is heterogeneous on such observational scales.  
The origin of the observed non-Gaussianity {\it might} be the above mentioned presence of big holes in the random
configuration of obstacles, that would lead to a finite probability of jumps much longer than the average,
corresponding to processes where collisions provide a nearly longitudinal
instantaneous velocity to the rod and at the same time the rod finds a long free ``corridor'' between the obstacles.
An investigation of the dynamics of the rod in an homogeneus liquid-like configuration of the obstacles,
where big holes are not present, could shed new light on this question. Work in this direction is in progress.

\begin{center}
ACKNOWLEDGEMENTS
\end{center}
\hspace{0.5 cm}
We are grateful to E. Frey for useful discussions. A.J.M. acknowledges
a postdoctoral grant form the Basque Government.

\begin {figure}
\begin{center}
\resizebox{0.92\columnwidth}{!}{
\includegraphics{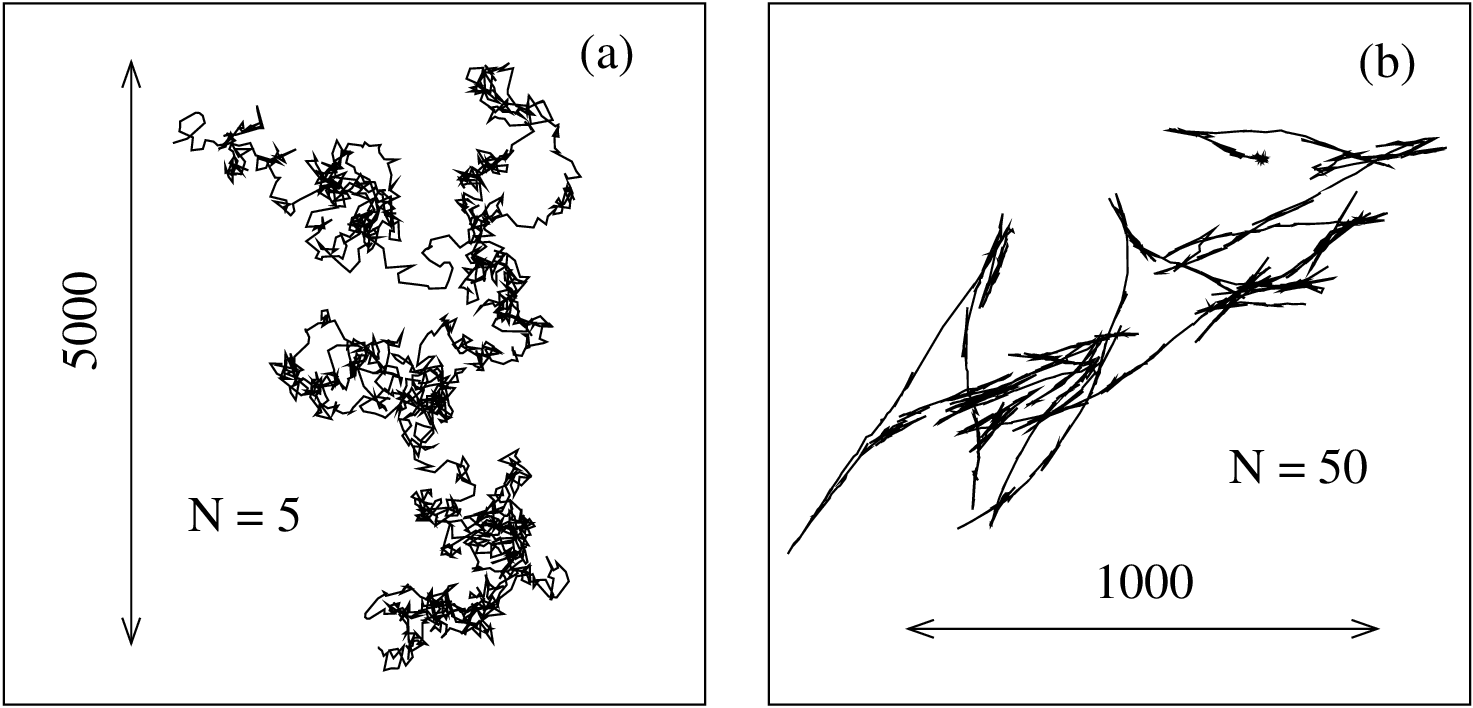}
}
\caption{Typical trajectories for $N=$ 5 and 50 for a complete simulation run of $t = 10^{6}$
($10^{8}$ time steps).}
\end{center}
\label{figure:1}
\end{figure}
\begin {figure}
\begin{center}
\resizebox{0.97\columnwidth}{!}{
\includegraphics{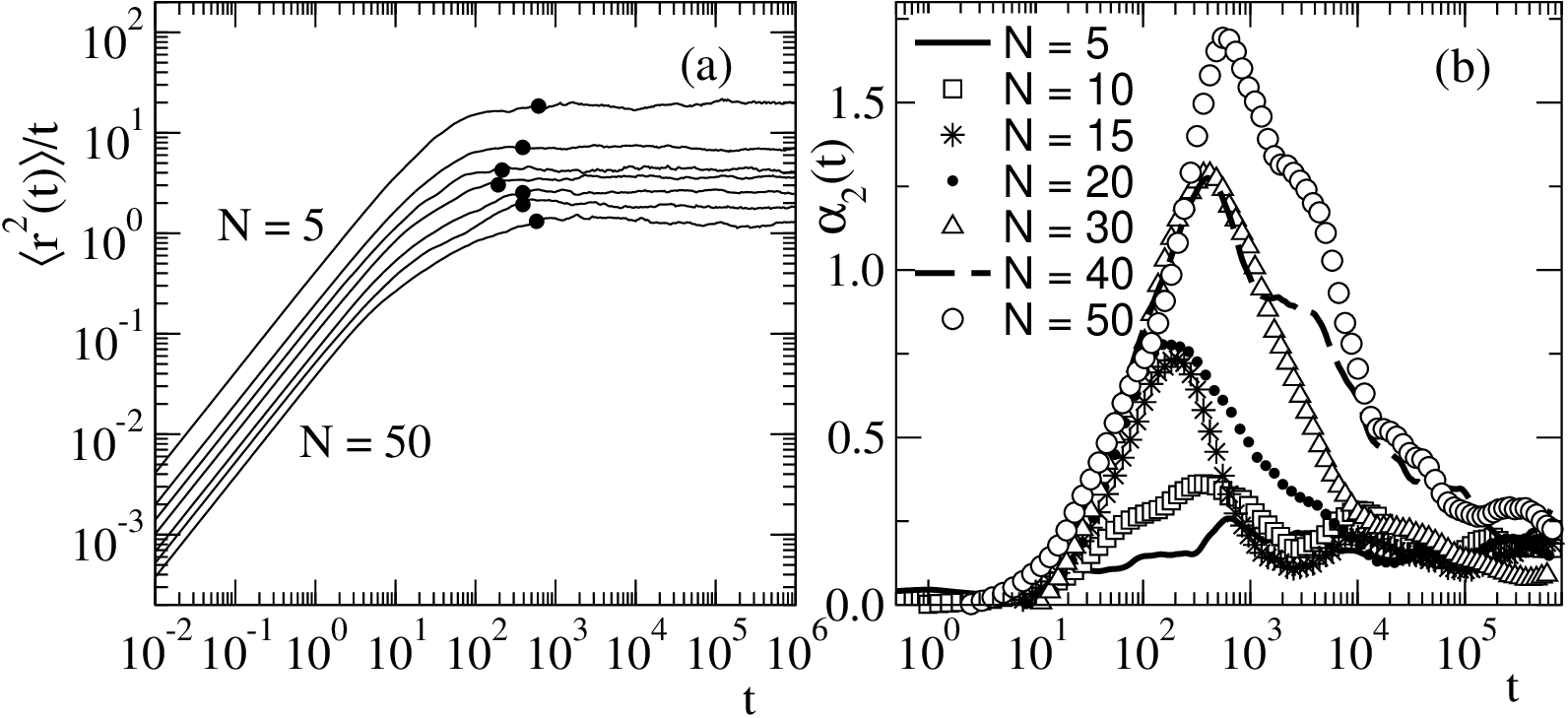}
}
\caption{(a): Mean-squared displacement of the center-of-mass divided by $t$ for
rod lengths $N =$ 5, 10, 15, 20, 30, 40, and 50 (top to bottom). Points mark
the time $t^{\ast}$ for the maximum of the non-Gaussian parameter.\newline
(b): Time-dependence of the non-Gaussian parameter for different rod lengths.}
\end{center}
\label{figure:2}
\end{figure}
\begin {figure}
\begin{center}
\resizebox{0.92\columnwidth}{!}{
\includegraphics{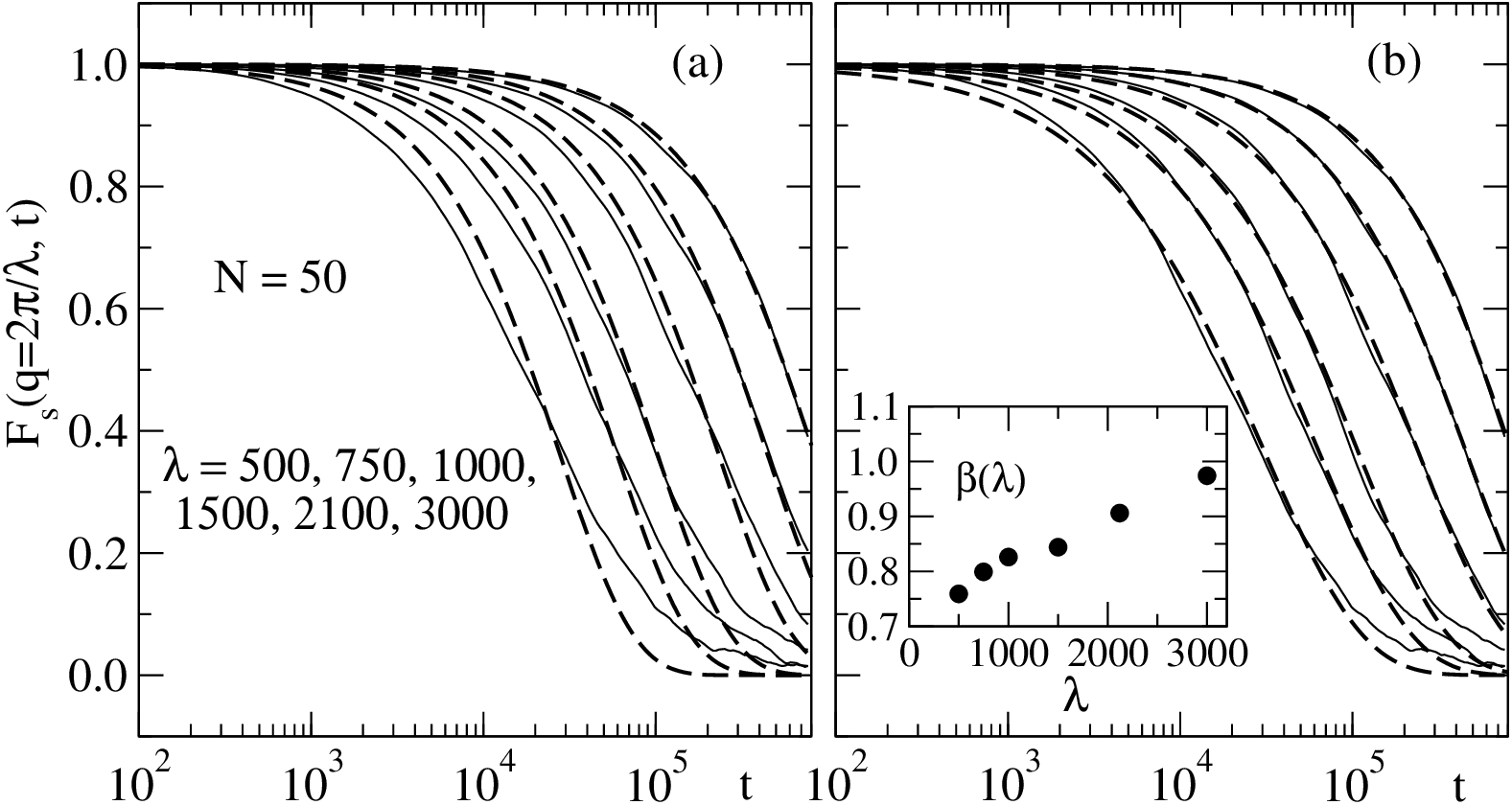}
}
\caption{Incoherent intermediate scattering function $F_{\rm s}(q,t)$ for $N = 50$ (solid lines) and
different values of $\lambda=2\pi/q$. Dashed lines are fits to an exponential (a) and to a KWW function
(b). The inset in (b) shows the dependence on $\lambda$ of the stretching exponent $\beta$.}
\end{center}
\label{figure:3}
\end{figure}
\begin{center}
REFERENCES
\end{center}
%
%
ALDER, B.J., and ALLEY, W.E., 1978, {\it J. Stat. Phys.}, {\bf 19,} 341.\\
BINDER, P.M., and FRENKEL, D., 1990, {\it Phys. Rev. A}, {\bf 42,} R2463.\\
BRUIN, C., 1972, {\it Phys. Rev. Lett.}, {\bf 29,} 1670.\\
CAPRION, D., and SCHOBER, H.R., 2000, {\it Phys. Rev. B}, {\bf 62,} 3709. \\ 
COLMENERO, J., ALVAREZ, F., and ARBE, A., 2002, {\it Phys. Rev. E}, {\bf 65,} 041804.\\
G\"{O}TZE, W., LEUTHEUSSER, E., and YIP, S., 1981a, {\it Phys. Rev. A}, {\bf 23,} 2634. \\
G\"{O}TZE, W., LEUTHEUSSER, E., and YIP, S., 1981b, {\it Phys. Rev. A}, {\bf 24,} 1008. \\
HANSEN, J.P., and McDONALD. I.R., 1986, {\it Theory of Simple Liquids}, (Academic Press London).\\
KOB, W., and ANDERSEN, H.C., 1995, {\it Phys. Rev. E}, {\bf 51,} 4626. \\
LORENTZ, H.A., 1905, {\it Arch. Neerl.}, {\bf 10,} 336. \\
MASTERS, A., and KEYES, T., 1982, {\it Phys. Rev. A}, {\bf 26,} 2129. \\
MOSSA, S., DI LEONARDO, R., RUOCCO, G., and SAMPOLI, M., 2000, {\it Phys. Rev. E}, {\bf 62,} 612. \\ 
PHILLIPS, J.C., 1996, {\it Rep. Prog. Phys.}, {\bf 59,} 1133. \\
PUERTAS, A.M., FUCHS, M., and CATES, M.E., 2002, cond-mat/0211087.\\ 
SCIORTINO, F., GALLO, P., TARTAGLIA, P., and CHEN, S.H., 1996, {\it Phys. Rev. E}, {\bf 54,} 6331.\\ 
VAN MEGEN, W., 2002, {\it J. Phys.: Condens. Matter}, {\bf 14,} 7699.\\
WEEKS, E.R., and WEITZ, D.A., 2002, {\it Chem. Phys.}, {\bf 284,} 361.\\
%
%
%
\end{document}